\documentclass[graybox]{svmult}

\usepackage{mathptmx}       % selects Times Roman as basic font
\usepackage{helvet}         % selects Helvetica as sans-serif font
\usepackage{courier}        % selects Courier as typewriter font
\usepackage{type1cm}        % activate if the above 3 fonts are
\usepackage{makeidx}         % allows index generation
\usepackage{graphicx}        % standard LaTeX graphics tool
\usepackage{multicol}        % used for the two-column index
\usepackage[bottom]{footmisc}% places footnotes at page bottom

\usepackage{color}        % color

\makeindex             % used for the subject index
\begin{document}

\title*{Fractal Reconnection in Solar and Stellar Environments}
\author{Kazunari Shibata and Shinsuke Takasao}
\institute{Kazunari Shibata \at Kwasan and Hida
Observatories, Kyoto University, Yamashina, Kyoto
607-8471\email{shibata@kwasan.kyoto-u.ac.jp}
\and Shinsuke Takasao \at Kwasan and Hida
Observatories, Kyoto University, Yamashina, Kyoto
607-8471 \email{takasao@kwasan.kyoto-u.ac.jp}}
\maketitle

\abstract{Recent space based observations of the Sun revealed that magnetic reconnection is ubiquitous in the solar atmosphere, ranging from small scale reconnection (observed as nanoflares) to large scale one (observed as long duration flares or giant arcades). Often the magnetic reconnection events are associated with mass ejections or jets, which seem to be closely related to multiple plasmoid ejections from fractal current sheet. The bursty radio and hard X-ray emissions from flares also suggest the fractal reconnection and associated particle acceleration. We shall discuss recent observations and theories related to the plasmoid-induced-reconnection and the fractal reconnection in solar flares, and their implication to reconnection physics and particle acceleration.  Recent findings of many superflares on solar type stars that has extended the applicability of the fractal reconnection model of solar flares to much a wider parameter space suitable for stellar flares are also discussed. 
}

\section{Introduction}
\label{sec:1}

The recent progress of space based solar observations in last few decades such as Yohkoh (1991-2001), SOHO (1995-) , TRACE (1998-2010), RHESSI (2002-), Hinode (2006-), SDO (2010-) has revolutionized the field of solar physics significantly.
With the help of space missions, it has been revealed that the solar corona is much more dynamic than had been thought, the quiet Sun is never quiet, the solar atmosphere is full of dynamic phenomena such as nanoflares, jets, waves, and shocks. 
An understanding that has emerged from the new observations is that {\it the magnetic reconnection is ubiquitous in the solar atmosphere}. 
So far, many pieces of evidence of magnetic reconnection have been found in solar flares and flare-like phenomena, and now we can say that the magnetic reconnection mechanism of solar flares is established,  at least, phenomenologically (see a review by Shibata and Magara 2011 \cite{Shibata2011}), although there are a number of unsolved problems that exist and these problems are highlighted in the present article. 
The long-standing puzzle of solar coronal heating mechanism has not yet been solved, although some of the new observations suggest that even quiet corona may be heated by small scale reconnection such as microflares, nanoflares, or picoflares (e.g., Parker 1988 \cite{Parker1988}, Priest and Forbes 2000 \cite{Priest2000}).

Virtually, almost all active phenomena occurring in the solar atmosphere seem to be related to magnetic reconnection, directly or indirectly.
This is probably a consequence of universal properties of magnetized plasmas: the solar corona is in a low plasma-$\beta (= p_{gas}/p_{mag} \ll 1)$ state,  where magnetic force and magnetic energy dominate over other force (e.g. gravitational forces) and energy, respectively. As a result, it is expected that the magnetic reconnection will have significant impact on heating as well as dynamics in the solar corona.
In addition, there is evidence that even the dynamic phenomena in the chromosphere (average $\beta \sim 1$) and photosphere (average $\beta \sim 10^4$) may be related to reconnection.
This is also a result of properties of magnetized plasma (e.g., 
Parker 1979, 1994 \cite{Parker1979},\cite{Parker1994}, Priest 1982 \cite{Priest1982}, Tajima and Shibata 1997 \cite{Tajima1997}):   
Magnetic fields tend to be concentrated to thin filaments in high $\beta$ plasmas, so that  the magnetic energy density in the filaments is much larger than the average value.
Therefore, once reconnection occurs in the filaments, the influence of reconnection can be significant.

On the other hand, some of the recent stellar observations have reported many dynamic activities in various stars such as jets and flares from young stellar objects and binary stars.  
 Even superflares have been discovered on many solar type stars. 
These dynamic events are much more energetic than solar flares, but the basic properties of these explosive events appear to be similar to the solar flares. 
Although evidence is still considered ''indirect", both theories and observations suggest similarity between solar flares and stellar flares. 

In this article, we provide a review on the recent observations of magnetic reconnection in solar flares and related phenomena in the solar atmosphere, with particular emphasis on a unified model of solar flares and flare-like phenomena based on the physics of magnetic reconnection.
The recent observations of stellar flares will also be discussed briefly.

\section{Fundamental Problems with Magnetic Reconnection in Solar Atmosphere}

There are some fundamental puzzles that need to be solved 
in order to fully understand the physics of solar and stellar flares. 

First, we have to deal with the most basic problem related with magnetic reconnection:

(1)  What determines the Reconnection Rate ?

 Recent magnetospheric observations and collisionless plasma theory suggest that fast reconnection (defined as the reconnection with the rate nearly independent of the Lundquist number) occurs if the current sheet thickness becomes comparable to ion Larmor radius ($r_{Li}$) and ion inertial length ($\lambda_i$) 
(either with anomalous resistivity or collisionless conductivity, 
see review by  e.g., Yamada et al. 2010 \cite{Yamada2010}):
$$  r_{Li} = {c \over eB} (m_i kT)^{1/2}
  \simeq 100 \Bigl({T \over 10^6 \ {\rm K}} \Bigr)^{1/2}
            \Bigl({B \over 10 \ {\rm G} }\Bigr)^{-1}
                         \ \  {\rm cm},  \eqno(1)$$
$$  \lambda_i = {c \over \omega_{pi}} 
  \simeq 300 \Bigl({n \over 10^{10} \ {\rm cm}^{-3}} \Bigr)^{-1/2}
                         \ \  {\rm cm},  \eqno(2)$$
where $\omega_{pi}$ is the ion plasma frequency.

However, the typical size of solar flares ($L_{flare}$) is 
$$  L_{flare} \simeq 10^9 - 10^{10} \ \ {\rm cm},  $$ 
and is much larger than the micro-plasma scales. 

Such enormous gap between micro- and macro- scales (ratio of both scales 
$\sim 10^7$) in solar flares is quite different from 
the situation of plasmas in magnetospheric and laboratory plasmas
where both scales are not so different, only within a factor of 100
(Terasawa et al. 2000 \cite{Terasawa2000}). 

Hence for the solar (as well as stellar) reconnection problem, 
one has to solve the following additional fundamental problem:

(2) How can we reach such a small scale to switch on anomalous resistivity or collisionless reconnection in solar flares ?

Finally, nonthermal emissions are one of the most important 
characteristics of the solar and stellar flares (and also of
other astrophysical flares and bursts). 
The nonthermal emissions are a result of acceleration of
electrons (10 keV - 1 MeV) and ions (10 MeV - 1 GeV).
However, not only the acceleration mechanism but also
the acceleration site have not yet been understood very well
(see review by Miller et al. 1997 \cite{Miller1997}
and Aschwanden 2002 \cite{Aschwanden2002}).

(3) What is the acceleration mechanism of high energy
particles in solar flares and what is the relation to
reconnection ?

In this article, we would argue that the aforementioned fundamental
puzzles are closely related each other and that
plasmoid-induced-reconnection process occurring in the current sheet and fractal reconnection are the key to the fundamental problems related with the magnetic reconnection in the solar and stellar atmosphere.

\section{A Unified View of Solar Flares and Flare-like Phenomena 
in the Solar Atmopshere}
\label{sec:2}

\subsection{Solar Flares,  Coronal Mass Ejections, and Plasmoid Ejections}

Solar flares have been observed with H$\alpha$ line from the ground based observatories, and are known to show two ribbon bright patterns in H$\alpha$ images. 
Motivated by the observations, a standard magnetic reconnection model called {\it CSHKP model} (after Carmichael 1964 \cite{Carmichael1964}, Sturrock 1966 \cite{Sturrock1966}, Hirayama 1974 \cite{Hirayama1974}, Kopp and Pneuman 1976 \cite{Kopp1976}) has been proposed. 
The CSHKP model predicts the formation of hot, cusp-shaped flare loops or arcades. 
The predicted cusp-shaped flare loops were indeed discovered by Yohkoh soft X-ray observations
(Tsuneta et al. 1992 \cite{Tsuneta1992}, Tsuneta 1996 \cite{Tsuneta1996}). 
Now, the standard reconnection model (CSHKP) of solar flares and flare-like phenomenon is considered well established.

However, cusp-shaped flares are rather rare, and many flares do not show 
clear cusps. 
Observations show that the shape of cusp in Soft X-rays is clear mainly during the {\it long duration event (LDE) flares}, that are long lived (more than 1 hours) flares,  large in size, but have small frequency of occurrence. 
On the other hand,  many flares (often called {\it impulsive flares})  are short lived, small in size, with large occurrence frequency, but show only a simple loop structure.
Therefore people sometimes argued that the observed ``simple loop'' structure of many flares is anti-evidence of magnetic reconnection. 

It was Masuda in 1994 \cite{Masuda1994} who changed the entire scenario. 
He discovered the {\it loop top hard X-ray source} well above the simple soft X-ray loop. Since hard X-ray source is produced by high energy electrons, it provided an important evidence that a high energy process related to the central engine of flares is occurring {\it not} in the soft X-ray loop but above the loop.  Hence even non-cusped loop flares may be energized by the magnetic reconnection high above the loop in a similar way as the reconnection in the cusp-shaped flares (Masuda et al. 1994 \cite{Masuda1994}). 
Since then, a unified model has been proposed in which the plasmoid ejection well above the loop top hard X-ray source is considered
(Shibata et al. 1995 \cite{Shibata1995}). 

%  Fig 1
\begin{figure}[t]
\sidecaption[t]
 \includegraphics[scale=.4]{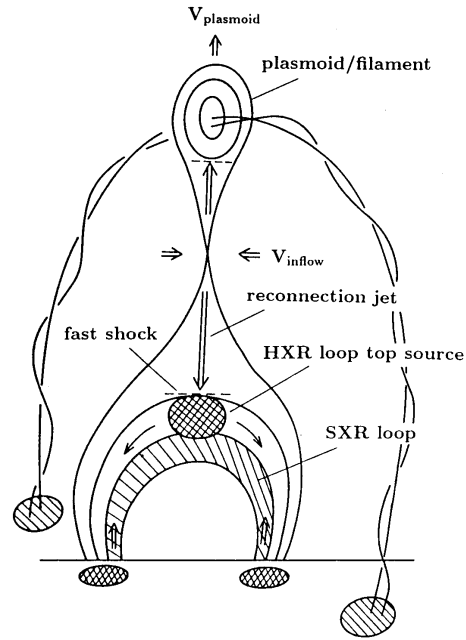}
  \caption{A unified model ({\it plasmoid-induced-reconnection model}) of solar flares and flare-like phenomena (Shibata et al. 1995)\cite{Shibata1995}, where LDE flares (Tsuneta et al. 1992 \cite{Tsuneta1992}) and impulsive flares are unified (Masuda et al. 1994 \cite{Masuda1994}).  
}
\label{fig:1}       % Give a unique label
\end{figure}

Indeed, many plasmoid ejections have been discovered above the Masuda-type flare loop (Shibata et al. 1995 \cite{Shibata1995}, Tsuneta 1997 \cite{Tsuneta1997},  Ohyama and Shibata 1997 \cite{Ohyama1997}, Ohyama and Shibata 1998 \cite{Ohyama1998}, Ohyama and Shibata 2000 \cite{Ohyama2000},
Shimizu et al. 2008 \cite{Shimizu2008}).  
It is important to note that the strong acceleration of plasmoid occurs during the impulsive phase of the flares. 
This may provide a hint to understand why and how a fast reconnection occurs in actual flares (Shibata and Tanuma 2001)\cite{Shibata2001}. 

About the half of the observed coronal mass ejections (CMEs) occur in association with flares, but the other half are not associated with flares.  
This also led to a lot of confusion in the community because CMEs were thought to be fundamentally different from flares.
However, Yohkoh/SXT revealed the formation of {\it giant arcade} at the 
feet of CMEs. 
These giant arcades are very  similar to cusp-shaped flares in morphology, but very faint in soft X-rays and H$\alpha$, and cannot be seen in non-imaging observations of soft X-rays (such as GOES) or hard X-rays. 
Only high-sensitive soft X-ray imaging observations were able to reveal the existence of giant arcade and the association of most of the non-flare CMEs with giant arcades.

\subsection{Microflares, Nanoflares, and Jets}

Space based solar observations revealed that the solar atmosphere is full of small scale flares, called microflares, nanoflares, and even picoflares,  and that these small scale flares are often associated with jets.  
One of the nice example of a  jet is X-ray jets discovered by Yohkoh/SXT 
(Shibata et al. 1992 \cite{Shibata1992}, Shimojo et al. 1996 \cite{Shimojo1996}). 
There are many pieces of observational evidence that shows that the jets are produced by magnetic reconnection (Shibata 1999)\cite{Shibata1999}. 
Yokoyama and Shibata (1995, 1996)\cite{Yokoyama1995} ,
 \cite{Yokoyama1996} performed MHD simulation of reconnection between an emerging flux and an overlying coronal field and successfully explained the observational characteristics of X-ray jets on the basis of their simulation results.
A direct extension of the 2D model to 3D MHD simulation has been carried out by 
Isobe et al. (2005, 2006)\cite{Isobe2005} \cite{Isobe2006}, where it was pointed out that the onset of the Rayleigh-Taylor instability at the top of the rising emerging flux leads to intermittent jets during reconnection. As a result, filamentary structures are formed naturally and are associated with patchy reconnection, that is in agreement with observations.
As for the more recent development of 3D models, see e.g.,  
Moreno-Insertis et al. (2008)
\cite{Moreno-Insertis2008} , Pariat et al (2010) \cite{Pariat2010}
Archontis and Hood (2013)
\cite{Archontis2013}.

%fig 2
\begin{figure}[t]
\sidecaption[t]
 \includegraphics[scale=.2]{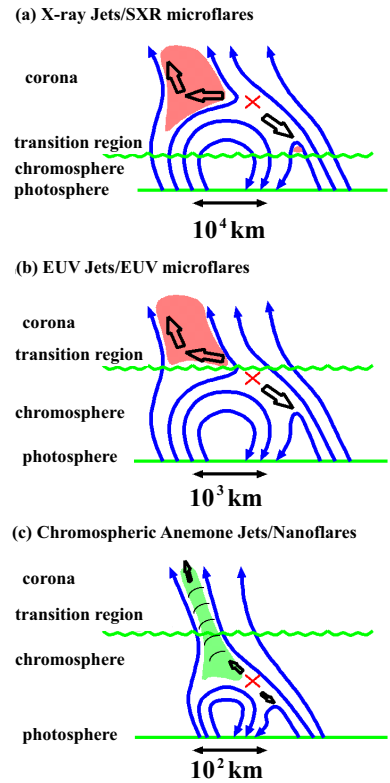}
  \caption{A schematic illustration of magnetic reconnection that occurs at various altitudes in the solar atmosphere (Shibata et al. 2007)\cite{Shibata2007}
}
\label{shibata2007}
\end{figure}

From the high resolution images taken with Hinode/SOT, 
Shibata et al. (2007)\cite{Shibata2007}  discovered 
numerous, tiny {\it chromospheric anemone jets} 
(whose apparent foot-point structures are similar to "sea anemone" in a three dimensional space)
 in the active region chromosphere.
The morphology of the chromospheric anemone jets is
quite similar to that of the coronal X-ray jets (Shibata et al. 1992
\cite{Shibata1992}, Shimojo et al. 1996\cite{Shimojo1996}, 
Cirtain et al. 2007\cite{Cirtain2007}), suggesting
that magnetic reconnection is occurring at the feet of these jets (Takasao et al. 2013 \cite{Takasao2013}),
although the length and velocity of these jets are much smaller
than those of the coronal jets (Table 1).

% section 3.3
\subsection{Unified Model : Plasmoid-Induced-Reconnection Model}

Table 1 summarizes solar ``flare'' observations from microflares to giant arcades.
 The size and time scales range in wide values,  from 200 km and 10 sec for nanoflares to $10^6$ km  and 2 days for  giant arcades.
However, it is interesting to note that if we normalize the time scale by the Alfven time, then the normalized time scale becomes similar,   $100 - 300 t_A$ (Alfven time). 
So the "flares" mentioned in Table 1 can be unified by a common physical process i.e. magnetic reconnection.
It is quite evident that although mass ejections are common in these "flares", the morphology is quite different between the large scale and small scale flares.
In large scale flares (e.g., giant arcades, LDE flares, impulsive flares), mass ejections  (CMEs, filament eruptions) are bubble like or flux rope type,  while in small scale flares (e.g., microflares, nanoflares), 
mass ejections are jets or jet-like.  
So what causes such morphological differences between "flares"?

Our answer to the question on morphology is as follows. According to our view (Fig. 3),  the plasmoid ejection is a key process that leads to a fast reconnection  (so we call ``plasmoid-induced-reconnection''), since plasmoids  (magnetic islands or helical flux ropes in 3D) are created naturally in the current sheets as a result of the tearing instability. 
In the case of large scale flares,  plasmoids (flux ropes) can retain their coherent structures during the ejection even during the interaction with the ambient magnetic field. 
Therefore many CMEs look like the flux rope ejection. However, in the case of small scale flares, plasmoids will lose their coherent shape soon after reconnection with the ambient field, and are likely to disappear (or lose their structure) eventually after the interaction (collision) with the ambient field. 
As the remnant (eventually), one would expect a spinning helical jet along the reconnected field lines along with generation of Alfv\'en waves. 
We conjecture that it will explain why jets are usually observed in association with small scale flares, although this idea should be tested through future observations. 
It is interesting to mention that some of the observations (Kurokawa et al. 1987\cite{Kurokawa1987}, Pike and Mason 1998\cite{Pike1998}, 
Alexander and Fletcher 1999\cite{Alexander1999}) have revealed the formation of spinning (helical) jets (Shibata and Uchida 1985 \cite{Shibata1985}) after flare-like phenomena.
Further, from the Hinode/XRT observations, Shimojo et al. (2007) \cite{Shimojo2007} found that an X-ray loop ejection (possibly helical loop ejection) finally led to an X-ray jet. These observations support the unified model shown in Fig. 3.

% fig 3
\begin{figure}[t]
\sidecaption[t]
 \includegraphics[scale=0.6]{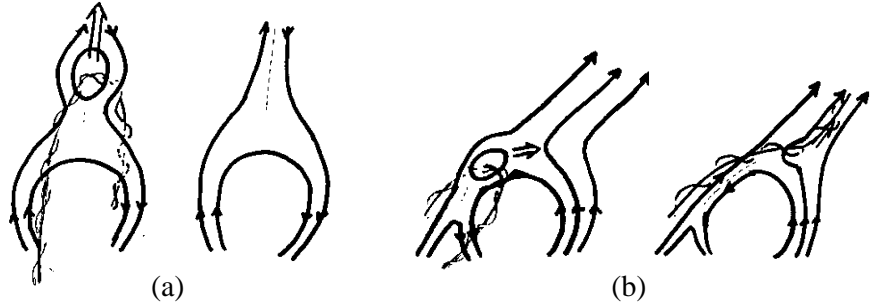}
  \caption{A unified model ({\it plasmoid-induced-reconnection model}) of solar flares and flare-like phenomena (Shibata 1999)\cite{Shibata1999}: (a) large scale flares (giant arcades, LDE flares, impulsive flares),  (b) small scale flares (microflares, nanoflares).}  
 \label{shibata1999}
\end{figure}

\begin{table} 
\caption{Summary of Observations of Various \lq \lq Flares''}
\label{tab:1}
\begin{tabular}{  l  l  l  l  l  l } \hline
\lq \lq flare'' &  length scale $(L)$ &  time scale $(t)$ &  Alfven time ($t_A$)
& $t/t_A$   & type of mass  ejection \\ 
     & ($10^4$ km) & (sec) &  (sec) &  & \\ \hline
     &             &       &        &  &  \\
%   $?$     &  $<$ 0.02  & $<$ 200  &  $<$ 20   &   \sim 10   & spicule  \\
%   &  &  &  &  \\
 nanoflares & $0.02 - 0.1$  &  20$-$100 &  $1-10$  & $10-50$  & chromospheric anemone jet \\
   &   &  &  &  \\    
microflares & $0.1 -  1$  &  100 $-$ 1000 &  $1-10$  & $\sim 100$  & coronal jet/surge \\
   &   &  &  &  \\
impulsive flares  & $1 - 3$ & 60 $- 3 \times 10^3$  &
        $10-30$   &  $60 - 100$ & plasmoid/filament   \\
     &   &  &  &  & eruption \\
LDE flares & $10 - 40 $  & $ 3 \times 10^3- 10^5$  & 
            $ 30 - 100 $  &  $100-300$ &  CME/plasmoid/  \\
     &   &  &  &   & filament eruption \\
giant arcades   & $30 - 100$  & $10^4 - 2 \times 10^5$  &
                $100-1000$  & $100-300$ &  CME/plasmoid/ \\
     &   &  &  &  & filament eruption \\
              \\ \hline
\end{tabular}
\end{table}

% section 4
\section{Plasmoid-Induced-Reconnection and Fractal Reconnection}

\subsection{Plasmoid-induced reconnection}
\label{section:plasmoid-induced-reconnection}

As we have discussed in the previous section, it has become clear that
the plasmoid ejections are observed quite often in solar flares 
and flare-like events.  As the spatial and temporal resolutions of 
the observations have become better, 
more and more, smaller plasmoids 
have been discovered in association with flares.  
So, how does plasmoid ejections in flares are related with the fast reconnection?

From the Soft and Hard X-ray observations of impulsive flares, Ohyama and Shibata (1997)  \cite{Ohyama1997} found that 
(1) a plasmoid was ejected long before the impulsive phase, 
(2) the plasmoid acceleration occurred during 
the impulsive phase (see Fig. 4(a)). 
As a result of the magnetic reconnection, plasmoid formation takes place (usually about 10 min) before the impulsive phase.  
When the fast reconnection ensues (i.e., in the impulsive phase), particle acceleration and huge amount of energy release occurs for $\sim10 t_A$.  
During this process the plasmoid acceleration is closely coupled to the reconnection inflow.

% Fig 4
\begin{figure}[t]
\sidecaption[t]
 \includegraphics[scale=.3]{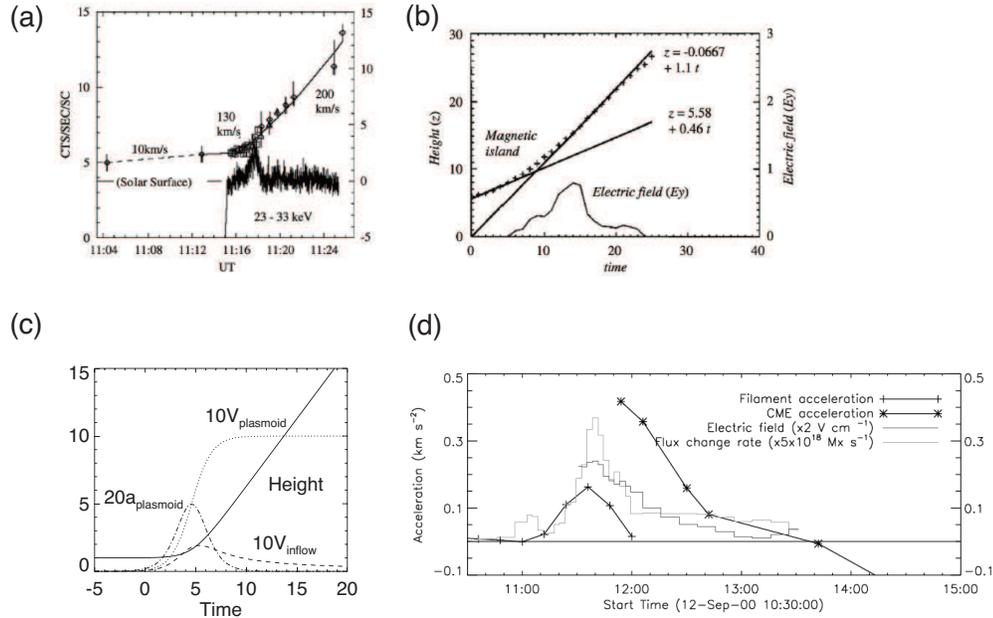}
    \caption{ (a)~Time variations of the height of an observed    plasmoid as well as hard X-ray intensity. From Ohyama and Shibata (1997)\cite{Ohyama1997}. (b)~Height-time relation of a magnetic island in a two-dimensional numerical simulation, which is supposed to be   the two-dimensional counterpart of a plasmoid. Time variation of    the electric field (i.e., the reconnection rate $\propto V_{inflow} $  is also plotted. From Magara et al. (1997)\cite{Magara1997}.  (c)~Analytical model of plasmoid acceleration in the plasmoid-induced-reconnection model.  From Shibata and Tanuma (2001)\cite{Shibata2001}, (d)~ Observations of a CME and associated filament eruption (Qiu et al. 2004)\cite{Qiu2004}. It is seen that the filament acceleration
(+) show the time variation similar to that of electric field (reconnection rate; thick solid curve). }\label{fig:4}
\end{figure}

A similar relation between the energy release (and fast reconnection) and
plasmoid acceleration has also been found in the case of CMEs 
(e.g., Zhang et al. 2001\cite{Zhang2001}, 
Qiu et al. 2004\cite{Qiu2004}; see Fig. 4(d))
as well as in laboratory experiment 
(Ono et al. 2011\cite{Ono2011}). 
What is the physical understanding that can be drawn from the relation between 
the plasmoid ejection and the fast reconnection ? 

It was Shibata and Tanuma (2001) \cite{Shibata2001} who suggested that 
plasmoid ejection induces a strong inflow
into the reconnection region 
as a result of mass conservation, and drive fast reconnection. 
Since the inflow (that determines the reconnection rate)
is induced by the plasmoid motion,
the reconnection process was termed as {\it plasmoid-induced reconnection}
(Shibata et al. 1995\cite{Shibata1995}, 
Shibata 1999\cite{Shibata1999}).

% Fig 5
\begin{figure}[t]
\sidecaption[t]
 \includegraphics[scale=.5]{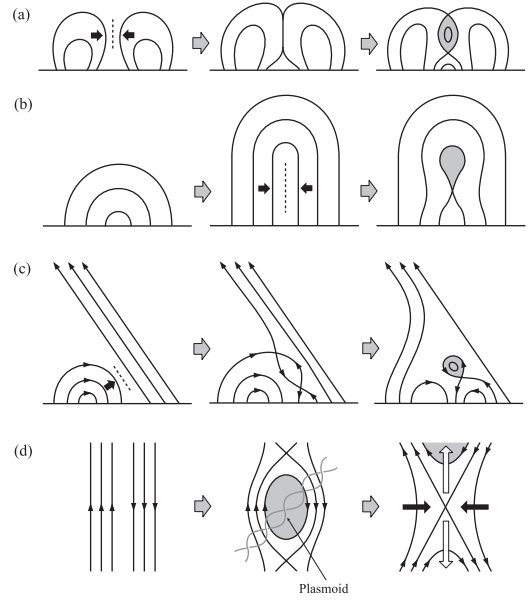}
    \caption{Schematic diagram of the plasmoid-induced reconnection model. The
solid lines indicate magnetic field lines. Panels (a)-(c) show the process creating
the plasmoid in the antiparallel magnetic field by the magnetic reconnection
in some typical magnetic field configurations.
Panel (d) shows how the plasmoid in the current sheet inhibits the 
reconnection, and how reconnection can 
occur, after the ejection of the plasmoid  (Nishida et al. 2009) \cite{Nishida2009}.  
}
    \label{fig:5}
\end{figure}

It should be noted that a plasmoid can be formed in any
current sheet (Fig. 5)  if the current sheet length 
is longer than the certain critical length scale. The critical length scale of the plasmoid instability comes from the physics of tearing mode instability (Furth, Killeen, Rosenbluth 1963\cite{Furth1963}).

During the initial stages of plasmoid formation, the plasmoid stays in the current sheet and during this stage, the plasmoid reduces the speed of reconnection significantly by inhibiting the reconnection inflow towards the reconnection region. 
Only when the plasmoid ejects out from the current sheet, a substantial amount of magnetic flux can come towards the reconnection region and trigger a magnetic reconnection. 
This facilitates the ejection of the plasmoid via strong reconnection outflow (reconnection jet), further that in turn enables new magnetic flux towards the current sheet.
The positive feedback between plasmoid ejection and reconnection inflow is established and fast reconnection continues, and eventually a plasmoid continues to eject from the current sheet with the Alfv\'en speed.

The 2D MHD numerical simulations (Magara et al. 1997 \cite{Magara1997},
Choe et al. 2000\cite{Choe2000}, 
Tanuma et al. 2001\cite{Tanuma2001}) showed such dynamics very well. 
Figure 4(b) shows a height-time plot from a two-dimensional MHD simulation
(Magara et al. 1997\cite{Magara1997}), in which magnetic reconnection produces an ejecting magnetic island (two-dimensional counterpart of a plasmoid).  
The time variation of the electric field  is also plotted in the height-time plot. 
It is found that the electric field, that is also a measure of reconnection inflow and reconnection rate, becomes large when the magnetic island (plasmoid) is accelerated. 

When comparing the MHD simulation and observations, it is assumed that the time variation of electric field in the reconnection region is closely related to the time variation of hard X-ray emissions because the electric field can accelerate particles which contribute to producing hard X-ray emissions. 
The comparison suggests that the plasmoid ejection drives a fast magnetic reconnection. 
More detailed  investigations of plasmoid ejection are given in 
Choe and Cheng (2000)\cite{Choe2000},  where multiple ejection of plasmoids and associated HXR bursts are discussed.

Shibata and Tanuma (2001)\cite{Shibata2001}  (Fig. 4c) 
developed a simple analytical model for the velocity of an ejecting plasmoid
by assuming (1) mass conservation between inflow and outflow
$V_p W_p = V_{inflow} L_{p}$, and (2) the plasmoid is accelerated
by the momentum added by the reconnection outflow
$\rho_p L_p W_p dV_p/dt = \rho_0 V_{inlow} L_{p} V_A $, 
where $V_p$ is the plasmoid velocity, $W_p$ the plasmoid width,
$L_p$ the plasmoid length, $V_{inflow}$ the inflow velocity,
$V_A$ the Alfven velocity, 
$\rho_p$ the plasmoid density, $\rho_0$ the density of ambient plasma.   
From these simple assumptions, 
they obtained the plasmoid velocity.
%
%\begin{equation}
$$
V_p=\ {\frac{{V}_{A}\exp\left({\omega t}\right)}{\exp\left({\omega t} 
\right)-1+{V}_{A}/{V}_{0}}}.
\eqno(3)
$$
%\label{vplm}
%\end{equation}
%
 In Equation~(3), $\omega$ represents the 
%acceleration
velocity growth rate of a plasmoid, defined as
%
%\begin{equation}
$$
\omega ={\frac{\rho_{0} {V}_{A}}{{\rho_p }{L}}}.   \eqno(4)
$$
%\label{omgplm}
%\end{equation}
%

The plasmoid velocity $V_p$, its acceleration ($a_p= dV_p/dt$),
inflow velocity $V_{inflow}$, and the height of the plasmoid
obtained from the analytical model (Shibata and Tanuma 2001)\cite{Shibata2001}
are plotted in Figure 4(c). 
It is interesting to note that
the acceleration and the inflow velocity 
(or reconnection rate) derived from this simple analytical
model agree well with the observations
 (Qiu et al. 2004 \cite{Qiu2004}, see
Fig. 4(d)) as well as the numerical simulation results
(Cheng et al. 2003)\cite{Cheng2003}. 

A detailed relation between the plasmoid velocity and the reconnection rate has been investigated by performing a series of numerical experiments 
(Nishida et al. 2009)\cite{Nishida2009}.
An extension to 3D has also been developed by Nishida et al. 
(2013)\cite{Nishida2013},
and it was eventually revealed that the formation of multiple flux ropes (helically twisted field lines) 
in a reconnecting current sheet plays an important role in enhancing the reconnection rate. 
These experiments show that the reconnection 
rate (inflow velocity) becomes larger when the plasmoid is accelerated
further by 3D effect (e.g., the kink instability) compared with 2D, 
whereas if the plasmoid velocity is decelerated, the reconnection
rate becomes smaller. 
When the reconnection is inhibited, the plasmoid
motion (or acceleration) is stopped (Fig. 5d).

% Fig 6
\begin{figure}[t]
\sidecaption[t]
\includegraphics[scale=.4]{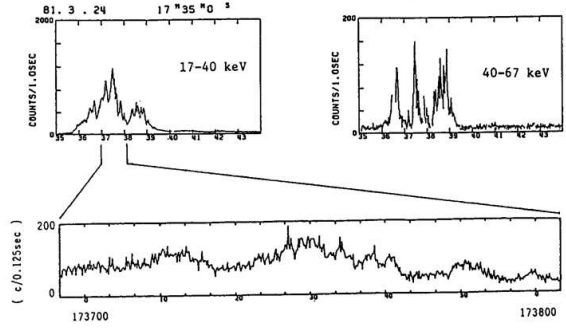}
  \caption{
Fractal-like time variability of
hard X-ray emission from a flare
 (from Ohki 1991 \cite{Ohki1991}).
}  
\label{fig:3}       % Give a unique label
\end{figure}

% section 4.2 
\subsection{Plasmoid Instability and Fractal Reconnection}

By performing 2D MHD simulation of the magnetic reconnection
on the current sheet triggered by a shock wave, 
Tanuma et al. (2001) \cite{Tanuma2001} found that 
(1) The reconnection does not start immediately after the passage
of the shock wave across the current sheet. Instead, the current 
sheet slowly change the shape as a result of the tearing
instability and becomes very thin in a fully nonlinear stage.
(2) The current-sheet thinning is saturated when the sheet thickness becomes comparable to that of the Sweet-Parker sheet. Then, Sweet-Parker reconnection starts, and the current-sheet length increases. 
(3) A secondary tearing instability occurs in the thin Sweet-Parker current sheet.
(4) As a result of the secondary tearing instability, further current-sheet thinning occurs.
(5) If the sheet becomes sufficiently thin to produce anomalous resistivity, 
a Petschek reconnection starts. 

On the basis of the nonlinear MHD simulations, 
Shibata and Tanuma (2001) \cite{Shibata2001} 
proposed that the current sheet 
eventually has a {\it fractal structure} consisting of many 
magnetic islands (plasmoids) with different sizes.

Once the current sheet has a fractal structure, 
it becomes possible to connect macro scale dynamics 
(with flare size of $10^9$ cm) and micro plasma scale dynamics
 (with ion Larmor radius or ion skin depth of $10^2$ cm).  
Then collisionless reconnection or anomalous resistivity
can be applied to flare reconnection problems
(see e.g., Cassak et al. 2005\cite{Cassak2005}, 
Daughton et al. 2009\cite{Daughton2009},
for the role of collisionless effects in reconnection).

The secondary instability of the Sweet-Parker
sheet has been discussed by Biskamp (1993, 2000) \cite{Biskamp1993}
\cite{Biskamp2000}. 
According to Biskamp (2000)\cite{Biskamp2000}, 
the condition of the secondary instability is that the tearing 
time scale  ($ t_{tearing} \sim  2 (t_A t_d)^{1/2} \sim 2 t_A {S_*}^{1/2}$, 
where $S_* = t_d/t_A = \delta V_A/\eta$ is the Lundquist number
with respect to the sheet thickness $\delta$) is shorter
 than the flow time scale ($t_{flow} \sim 0.5 L/V_A \sim 0.5 t_d$),
 \footnote{In the Sweet-Parker sheet, we find $L/V_A = \delta/V_{inflow} = \delta^2/\eta = t_d.$}
where $t_A = \delta/V_A$ is the Alfven time (across the sheet), 
$t_d = \delta^2/\eta$ is the diffusion time, 
$\delta$ is the thickness of the Sweet-Parker sheet, 
$L$ is the length of the sheet, 
$\eta$ is the magnetic diffusivity.
From these relations, we find that the condition of the
secondary instability is
$$  t_d^{-1}  <  0.25 (t_d t_A)^{-1/2}      \eqno(5)$$
 or  $  (t_d/t_A) > 16. $
Using the global Lundquist number $S = LV_A/\eta$,  we find
$t_d/t_A = (\delta/L)  (LV_A/\eta) = (\delta/L)  S 
= S^{1/2}$ 
for the Sweet Parker sheet ($\delta/L = S^{-1/2}$).
Then the above condition can be written
$ L/\delta > 16. $
For more accurate calculation for large Lundquist number, 
this condition becomes $L/\delta > 10^2$ (or $S > 10^4$) 
(Biskamp 2000)\cite{Biskamp2000}.  
This condition roughly explains the result of
Tanuma et al. (2001)\cite{Tanuma2001}.

Shibata and Tanuma (2001) \cite{Shibata2001} 
calculated how the current 
sheet becomes thinner as a result of the secondary tearing instability
(see Fig. 7a)
whose condition is given by
$$
{\delta_n \over L} \leq A \Bigl({\delta_{n-1} \over L } \Bigr)^{5/6}
\eqno(6)
$$
where $A = 6^{2/3} S^{-1/6}$ and $S=L V_A/\eta$. 
\footnote{
Here Shibata and Tanuma (2001) \cite{Shibata2001} 
assumed that
the condition of the secondary instability 
was  $ (t_A t_d)^{1/2} < L/V_A$. 
If we use more rigorous condition
 $ (t_A t_d)^{1/2} < a L/V_A$, where $a \simeq 4$
(see above), we find 
$A = (a b)^{2/3} S^{-1/6}$.
Note that Shibata and Tanuma (2001) 
assumed $a = 1$ and $b = 6$.
If we use $a = 4, b = 2 \pi = 6.28$, then 
$A \simeq 0.05$ for $R_m = 10^{13}$. 
In this case, we find $n = 12$ for
the condition that the sheet thickness
becomes less than the ion Larmor radius
($ \sim 100$ cm),  i.e., 
$\delta_n/L < 10^{-7}$ and the initial
sheet  thickness and length are $10^8$ cm
and $10^9$ cm. }

From this, they obtain the solution
$$ {\delta_n \over L} \leq 
A^{6(1-x)} \Bigl({\delta_0 \over L} \Bigr)^x 
   \eqno(7)   $$
where $x = (5/6)^n$.

% Fig 7
\begin{figure}[t]
\sidecaption[t]
\includegraphics[scale=.4]{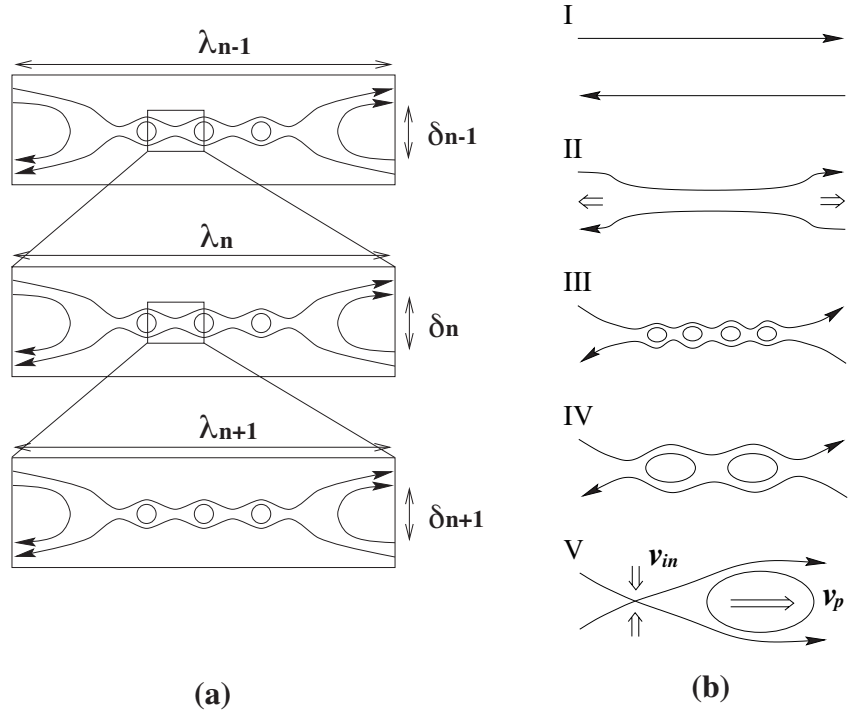}
  \caption{(a) Schematic view of the fractal reconnection. 
(b) A scenario for fast reconnection. 
I: The initial current sheet. 
II: The current sheet thinning in the nonlinear stage of the tearing instability
or global resistive MHD instability. The current sheet thinning stops
when the sheet evolves to the Sweet-Parker sheet. 
III: The secondary tearing in the Sweet-Parker sheet. 
The current sheet becomes fractal
because of further secondary tearing as shown in (a). 
IV: The magnetic islands coalesce with each other to form bigger 
magnetic islands.
The coalescence itself proceeds in a fractal manner. 
During the III and IV phases, a microscopic plasma scale 
(ion Larmor radius or ion inertial length) is reached, 
so that the fast reconnection becomes possible at small
scales, 
V: The greatest energy release occurs when the largest plasmoid
(magnetic island or flux rope) is ejected. The maximum inflow speed
($V_{inflow}$ = reconnection rate) is determined by the velocity of the plasmoid
($V_p$). 
Hence this reconnection is termed as {\it plasmoid-induced-reconnection}.
 (from Shibata and Tanuma 2001 \cite{Shibata2001}).
}  
\label{fig:7}       % Give a unique label
\end{figure}

Note that Shibata and Tanuma (2001)\cite{Shibata2001}
 did not assume that the sheet
is exactly the same as the Sweet-Parker sheet,
since in actual solar condition the Lundquist
number is so large that we cannot have the 
Sweet Parker sheet as the initial condition.
Instead they assumed that the sheet become 
unstable once the instability condition
$$ {\rm tearing \ \ time \ \ } (t_d t_A)^{1/2} = ({\delta_n}^3/(\eta V_A))^{1/2}
 < {\rm flow \ \ traveling  \ \ time \ \ } \lambda_n/V_A   $$  
is satisfied, where
 $\lambda_n$ is the most unstable wavelength and
 is given by $\lambda_n \simeq 6 \delta_n {S_{*,n}}^{1/4}$
 where $S_* = \delta_n V_A/\eta$.

For actual solar coronal condition, it is 
found $n \geq 6$ to reach microscopic scale such as
ion Larmor radius or ion skin depth 
$\delta_6/L < L_{ion-Larmor}/L \simeq 10^{-7}. $

Shibata and Tanuma (2001) presented a scenario for fast reconnection
in the solar corona as shown in Figure 7(b). 
That is, the current sheet becomes a fractal sheet consisting
of many plasmoids with different sizes. 
The plasmoids tend to coalesce with each other 
(Tajima et al. 1987 \cite{Tajima1987})
to form bigger plasmoids.  When the biggest island (i.e., 
monster plasmoid) is ejected out of the sheet, 
we have the most violent energy release which 
may correspond to the impulsive phase of flares.

Solar observations show
the fractal-like time variability of solar flare
emission, especially in microwaves 
(Karlicky et al. 1996\cite{Karlicky1996}, 
Aschwanden 2002\cite{Aschwanden2002}),  and hard X-rays 
(Ohki 1991 \cite{Ohki1991}; see Fig. 6).
The above idea of the fractal reconnection seems to explain
the observations very well, since the observations suggest
fragmented energy release processes in the 
fractal (turbulent) current sheet.  For example, Karlicky 
et al. (1996) \cite{Karlicky1996}
showed that the temporal power spectrum
analysis of the narrow band of dm-spikes 
of a flare show power-law spectrum,
suggesting Kolmogorov spectra after transformation
of the frequency scales to the distance scales. 

More recently, Singh et al. (2015)\cite{Singh2015} 
extended the fractal reconnection theory of Shibata and Tanuma to 
that in a partially ionized plasma in the solar chromosphere, and 
basically obtained the similar result.

It is interesting to note that 
Tajima and Shibata (1997) \cite{Tajima1997}
found the growth rate of the secondary tearing instability 
of the Sweet-Parker sheet has positive dependence on
the Lundquist number $\omega \propto S^{1/4}$ and
the most unstable wavelength decreases with
increasing with $S$ with the scaling $\lambda \propto S^{-3/8}$.

The tearing mode instability in Sweet-Parker current sheet is studied 
by Loureiro et al. (2007)\cite{Loureiro2007},
and the tearing instability of the Sweet-Parker sheet is now addressed as {\it plasmoid instability}.
Numerical simulations of the nonlinear evolution of
the plasmoid instability has been developed significantly 
in recent ten years, and will be discussed in detail in 
subsection 4.3.

% 4.3 
\subsection{Recent Development of Numerical Simulations of
Plasmoid-Dominated Reconnection}

\begin{figure}[t]
\sidecaption[t]
\includegraphics[scale=.3]{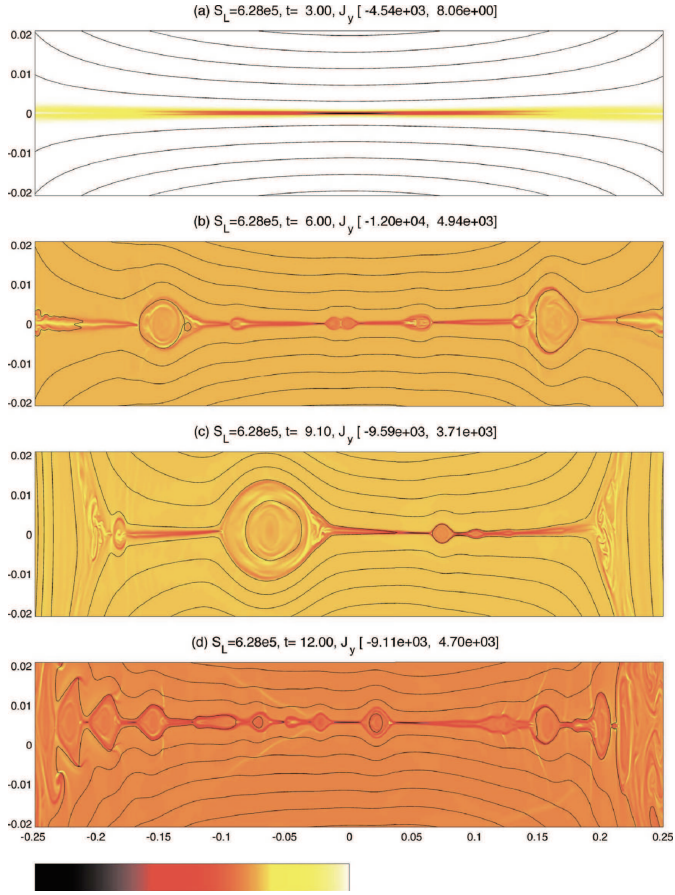}
\caption{Time-sequence of the nonlinear evolution of the current density $J_y$ of a Sweet-Parker current sheet in a large system of Lundquist number $S=6.28\times 10^5$. The black lines represent surfaces of constant $\psi$. (Bhattacharjee et al. 2009)}
\label{fig:Bhattacharjee2009_fig1}       % Give a unique label
\end{figure}

\begin{figure}[t]
\sidecaption[t]
\includegraphics[scale=.5]{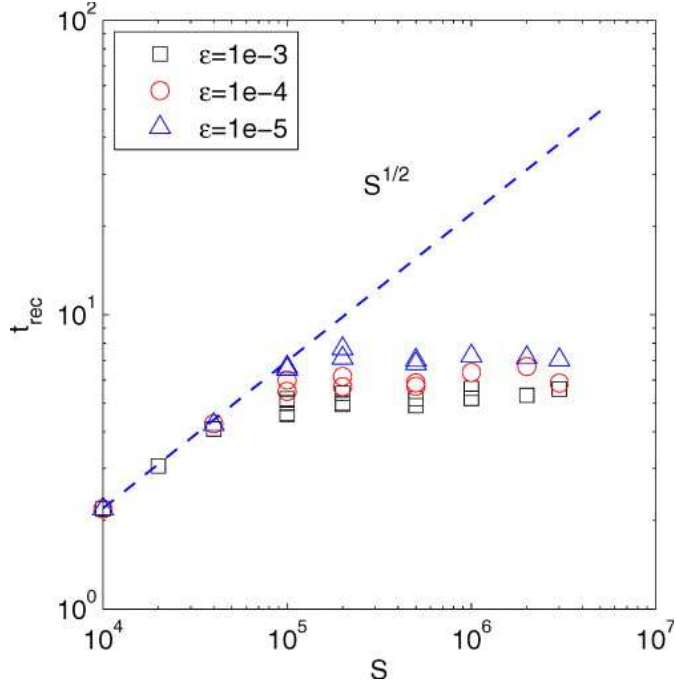}
\caption{The reconnection time $t_{\rm rec}$ for various $S$ and $\epsilon$. The dashed line is the Sweet-Parker scaling. (Huang \& Bhattacharjee 2010)}
\label{fig:Huang_Bhattacharjee2010_fig2}       % Give a unique label
\end{figure}

The nonlinear evolution of the plasmoid-dominated reconnection has been extensively investigated in recent years using MHD simulations. Samtaney et al. (2009)\cite{Samtaney2009} performed 2D MHD simulations of the formation of plasmoid chains in a very high-Lundquist number ($10^4<S<10^8$), and confirmed the scaling of the plasmoid number (or plasmoid distribution) in the linear regime ($\sim S^{3/8}$) predicted by Tajima \& Shibata (1997)\cite{Tajima1997} and Loureiro et al. (2007)\cite{Loureiro2007}. Cassak et al. (2009)\cite{Cassak2009}, Bhattacharjee et al. (2009)\cite{Bhattacharjee2009} and Huang \& Bhattacharjee (2010)\cite{Huang2010} found that once the plasmoid instability sets in, the reconnection rate becomes nearly independent of the Lundquist number (Figures~\ref{fig:Bhattacharjee2009_fig1} and \ref{fig:Huang_Bhattacharjee2010_fig2}). An energy cascade to smaller scales during a tearing process is clearly presented by Barta et al. (2011)\cite{Barta2011}. Since other studies have also confirmed this result (e.g. Loureiro et al. (2012)\cite{Loureiro2012}), it now seems to be a robust result. However, all the studies are restricted to 2D and to $S$ of $\sim10^8$. It is not obvious that the 2D results will remain unchanged for 3D astrophysical situations with a high-Lundquist number (e.g. $S\sim10^{13}$ for solar applications).

%Fig 8

% fig 9

The plasmoid distribution in the non-linear regime, which is essential for the understanding the current sheet thinning process, has been discussed by several authors. By considering a stochastic generation, growth, coalescence, and ejections of plasmoids, Uzdensky et al. (2010)\cite{Uzdensky2010} predicted the dependence of the plasmoid distribution function $f$ on flux $\Phi$ and plasmoid width $w_x$: $f(\Phi)\propto \Phi^{-2}$ and $f(w_x)\propto w_x^{-2}$ (a similar approach was independently done by Fermo et al. 2010\cite{Fermo2010}). Loureiro et al. (2012)\cite{Loureiro2012} performed 2D MHD simulations to investigate the plasmoid distribution, and obtained double-power-law-like distributions (Fig. 10). It was argued that the distribution with steeper power law at larger flux and width (large plasmoids) seems to scale as the relations by Uzdensky et al. (2010). Huang \& Bhattacharjee (2012)\cite{Huang2012} also studied the distribution, and found that the relative speed of plasmoids should be considered to understand the evolution of plasmoids. Considering this, a simple governing equation was constructed for the distribution function that gives the scaling $f(\Phi)\sim \Phi^{-1}$. The power law distribution has been confirmed by the following study by Huang \& Bhattacharjee (2013) . We note that the scaling $f(\Phi)\sim \Phi^{-1}$ can also be seen in the case of Loureiro et al. (2012). Observational tests for the scaling have just started (Guo et al. (2014)\cite{Guo2014}).

Considering the plasmoid-induced-reconnection scenario, emergence and ejections of large plasmoids from the current sheet play an important role in enhancing the reconnection rate and carrying a large amount of magnetic flux towards the reconnection regions. Emergence of abnormally large (with the size of $\sim$0.1 times the system size) "monster" plasmoids during a stochastic plasmoid-dominated reconnection was predicted by Uzdensky et al. (2010)\cite{Uzdensky2010}. Loureiro et al. (2012)\cite{Loureiro2012} studied the distributions of the magnetic flux of plasmoids and of the half-width of plasmoids, and found that monster plasmoids occasionally occur. 

%Fig 10
\begin{figure}[t]
\sidecaption[t]
\includegraphics[scale=.3]{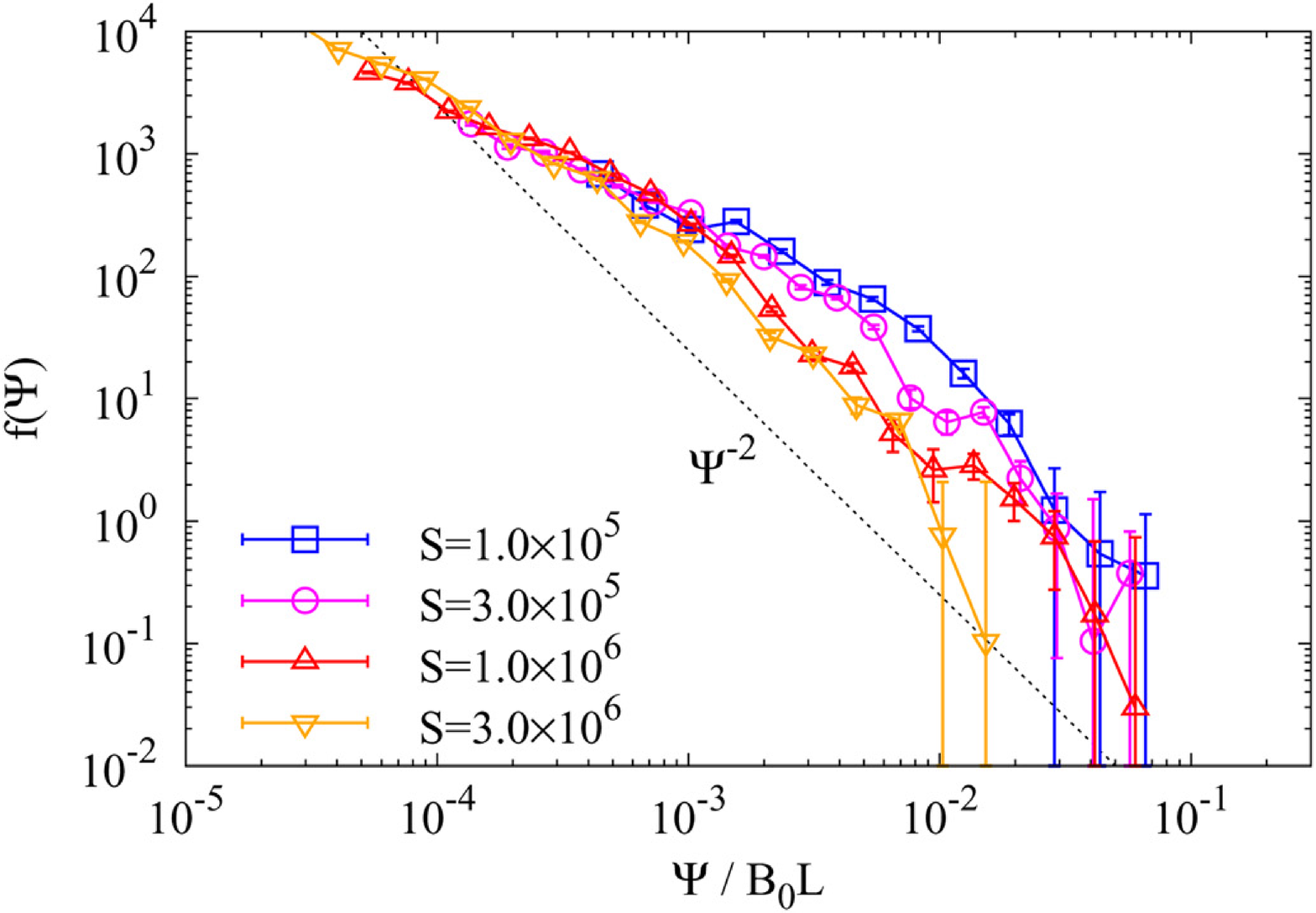}
\caption{Plasmoid distribution functions from direct numerical simulations. 
(Loureiro et al.  2012)}
\label{fig: Loureiro2012_fig3_left}       % Give a unique label
\end{figure}

Thanks to the modern computational resources, it has become possible to investigate the plasmoid-dominated reconnection in 3D. The first 3D simulation was presented by Linton \& Priest (2002)\cite{Linton2002}, in which a pair of perpendicular, untwisted magnetic flux tubes collide to form a current sheet. Although the spatial resolution was not enough to discuss the evolution of the reconnection rate, they found the formation and coalescence of flux ropes (corresponding to plasmoids in 3D). Wyper \& Pontin (2014)\cite{Wyper2014} for the first time studied non-linear plasmoid instability of 3D null point current sheets. Comparing a 2D plasmoid-dominated reconnection scenario, they found that (1) 3D current sheets are subject to an instability analogous to the plasmoid instability, but are marginally more stable than equivalent 2D neutral sheets, (2) an efficient 3D flux mixing leads to a substantial increase in the reconnection rate, and (3) the interaction of flux ropes appear to be driven primarily by kink instability which is a 3D instability.

The evolution of plasmoid chains in a relativistic Poynting-dominated plasma
 ($S=10^3-10^5$) was investigated by 
Takamoto (2013)\cite{Takamoto2013}, where the reconnection rate becomes nearly independent of the Lundquist number after the generation of plasmoids, similar to non-relativistic cases. This study indicates that the plasmoid formation plays an important role in fast reconnection even in relativistic plasma. 

% Fig 11
\begin{figure}[t]
\sidecaption[t]
 \includegraphics[scale=.4]{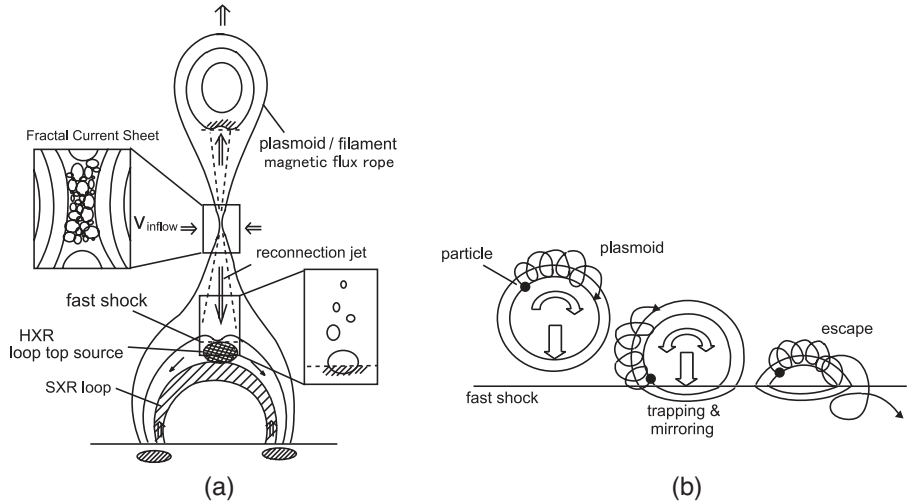}
  \caption{
Overall picture of a new particle acceleration mechanism in plasmoid-shock interaction. (a) Multiple plasmoids of various scales are intermittently ejected upward and downward out of a turbulent current sheet and collide with the termination shocks of reconnection outflows. (b) Scenario of shock acceleration at the fast shock trapped
in a plasmoid (Nishizuka and Shibata 2013)\cite{Nishizuka2013}. 
}  
\label{fig:12}
\end{figure}

The formation of plasmoids could be a key to understand the origin of energetic nonthermal particles. 
Drake et al. (2006) \cite{Drake2006} have pointed out that the contracting
plasmoids (magnetic islands) can accelerate electrons during reconnection 
because of Fermi-type processes that occur for electrons trapped in the contracting
magnetic islands. 
Nishizuka \& Shibata (2013)\cite{Nishizuka2013}) proposed when plasmoids
pass through the fast mode termination shock in the reconnection region, 
particles trapped in plasmoids can be accelerated via Fermi-type process. 
Namely, particles in a plasmoid are reflected upstream the shock front by magnetic mirror effect. 
As the plasmoid passes through the shock front, the reflection distance becomes shorter and shorter driving Fermi acceleration, 
until it becomes particle's Larmor radius (Fig. 11). 
The fractal distribution of plasmoids may also 
have a role in naturally explaining the power-law spectrum in nonthermal emissions.

Recently, much attention has been paid to plasmoid-dominated reconnection in a partially ionized plasma. Since the electron-ion collisional timescale is much shorter than most of the timescales of interest, neutral-ion two fluid effects have been extensively concerned. 
Ni et al. (2015)\cite{Ni2015} performed 2D MHD simulations with the effects of the ambipolar diffusion and radiative cooling to study the nature of reconnection in the solar chromosphere, where the ambipolar diffusion is a resistivity diffusion introduced by ion-neutral collisions (equivalently, a Pedersen resistivity). They investigated the role of both effects on the plasmoid instability changes in the presence of a guide field. They found that a fast reconnection takes place as a result of the plasmoid formation for zero as well as for strong guide field. When the current sheet becomes thin, the ion-neutral collisional timescale can be comparable to or shorter than a dynamical timescale, resulting in the decoupling of the neutral and ion fluids. In addition, ionization, recombination, and charge exchange processes will change the ionization degree depending on the local temperature and density, which will affect the removal processes of the neutrals and ions from the current sheet. Some multi-fluid treatments with the effects of ionization, recombination, and charge exchange are required to study the two fluid and non-equilibria partial ionization effects on the reconnection structure. Leake et al. (2013)\cite{Leake2013} performed two-fluid MHD simulations with the non-equilibrium partial ionization effects, and found a fast reconnection rate independent of the Lundquist number. In addition, it was found that the non-equilibrium partial ionization effects lead to the onset of the nonlinear secondary tearing instability at comparatively lower values of the Lundquist number than that has been reported in the case of fully ionized plasmas.

% Fig 12
\begin{figure}[t]
\sidecaption[t]
 \includegraphics[scale=.3]{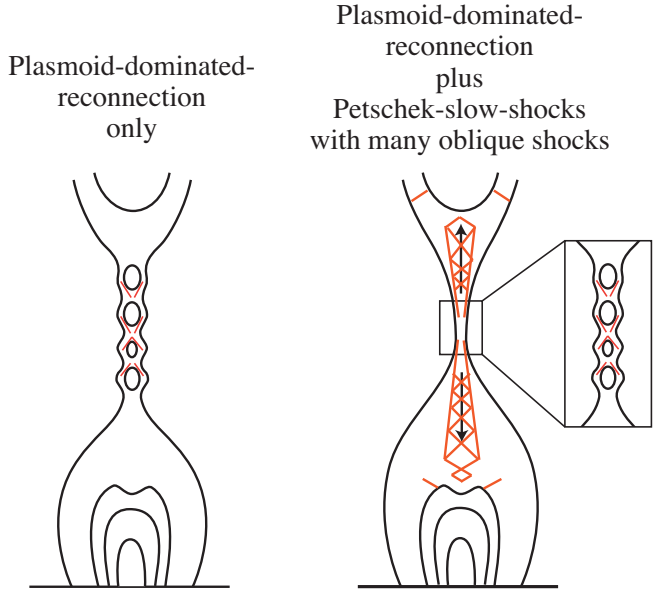}
  \caption{
plasmoid-dominated current sheet
vs plasmoid+Petschek slow shock
}  
\label{fig:12}
\end{figure}

Because shocks are crucial for the energy conversion process during the reconnection, the shock structure in and around plasmoids has been studied by many authors. It has been argued for a long time that slow shocks emanating from reconnection points (so-called Petschek-type slow shocks) cannot be established with a uniform resistivity. Tanuma et al. (2001)\cite{Tanuma2001} pointed out for the first time that Petschek-like slow shocks can emanate from an X-point in a tearing current sheet. However, due to the formulation of their resistivity model, it was not clear what is the origin of the formation of slow shocks:  due to the plasmoid nature, or due to the onset of the anomalous resistivity in their simulations. Recently, Mei et al. (2012)\cite{Mei2012} studied the evolution of the current sheet formed below the erupting CME using a uniform resistivity. They found that plasmoids are actually accompanied by Petschek-like slow shocks. Although it was not explicitly mentioned, the structure of the simulated current sheet seems to be a combination of plasmoid-dominated reconnection and global Petschek-like slow shocks. This motivates us to present a new view of flare reconnection shown in Figure 12. For the understanding of the shock structure of flaring regions, further studies are necessary. As for a variety of shock and discontinuity structure in and around a plasmoid, the reader is also referred to Zenitani \& Miyoshi (2011)\cite{Zenitani2011} and Zenitani (2015)\cite{Zenitani2015}.

% Fig 13
\begin{figure}[t]
\sidecaption[t]
\includegraphics[scale=.3]{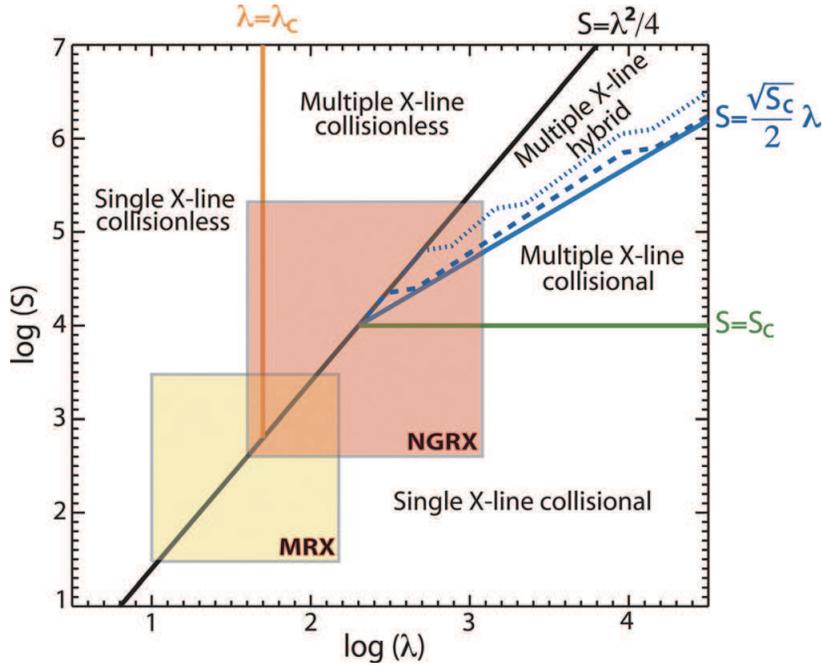}
\caption{A phase diagram diagram for magnetic reconnection in two dimensions. $\lambda$ and $S$ are th effective plasma size normalized by the ion skin depth and the Lundquist number of the system [From Ji \& Daughton 2011].}
\label{fig:Ji_Daughton2011_fig2}       % Give a unique label
\end{figure}

It would be worth noting that a fast reconnection can be obtained even in the MHD regimes if the Lundquist number is high enough to trigger the plasmoid instability. But it is the microphysics that may become important during the reconnection process. The recurrent plasmoid formation and ejection from the current sheet at multi-scales can lead to the formation of thin current sheets with the width of a microscopic scale like the ion skin depth or ion Larmor radius, and therefore some microphysics (e.g. anomalous resistivity) can set in at some time. At this stage, one would expect that microphysics will play a crucial role in determining the reconnection process: from {\it collisional} physics to {\it collisionless} physics. For more information on the various regimes that lie between collisional and collisionless processes, readers are referred to the discussion of Ji \& Daughton (2011)\cite{Ji2011} (Fig.~\ref{fig:Ji_Daughton2011_fig2}). The link between micro- and macro- scales should be explored in more detail.

% section 4.4
\subsection{Observational Evidence of Plasmoid-Dominated Reconnection
and Fractal Reconnection}

Asai et al. (2004) \cite{Asai2004} reported that there are multiple 
downflow (supra arcade downflow; McKenzie and Hudson 1999
\cite{McKenzie1999}, McKenzie et al.  2013\cite{McKenzie2013})  
which are associated with
hard X-ray impulsive emssions. Although the origin of 
supra arcade downflow is still not yet understood well,  
the physical relation between downflow and hard X-ray emission
may be similar to the relation  between plasmoid ejections and 
hard X-ray emissions (see Fig. 4a).

Using the data on post-CME current sheets observed by SOHO/UVCS,
Bemporad (2008) \cite{Bemporad2008}
examined the evolution of turbulence by interpreting 
the nonthermal broadening of the [Fe xviii] line profiles, and found 
that the turbulent speeds decay from 60 km/s to 30 km/s 
during 2 days after CME ejection.

Nishizuka et al. (2009) \cite{Nishizuka2009} examined the time variation of the
intensity of the flare kernels 
and found that intermittent radio/HXR bursts, whose peak intensity, duration, and time interval were well described by power-law distribution functions. 
This result may be evidence either of “self-organized criticality" in
avalanching behavior in a single flare event, or fractal current sheets in the impulsive reconnection region.

By analyzing the soft X-ray images and hard X-ray emission of a flare
taken with Yohkoh satellite, Nishizuka et al. (2010) \cite{Nishizuka2010}
found multiple 
plasmoid ejections with velocities of 250 - 1500 km/s.
They also found that each plasmoid ejection is associated with an impulsive
burst of hard X-ray emssions which are  a result of  high energy electron
acceleration and are signature of main energy release due to 
the fast reconnection. 

Singh et al. (2012) \cite{Singh2012}
analyzed chromospheric anemone jets (Shibata et al. 
2007\cite{Shibata2007})  observed by Hinode/SOT, and found that all the jets they analyzed
show intermittent and recurrent ejections of the jet and 
the corresponding brightening of the loop. 
Such behavior is quite similar to
plasmoid ejections from large flares (e.g., Nishizuka et al. 2010\cite{Nishizuka2010}). 
Note that chromospheric jets are considered to 
be a result of {\it collisional} magnetic reconnection 
in a weakly ionized plasma ({Singh et al. 2011 \cite{Singh2011}) . Nevertheless,
the time-dependent behavior of chromospheric jets is quite similar to that of
coronal reconnection ({\it collisionless} reconnection), 
suggesting the common macro-scale dynamics, i.e.,
plasmoid-induced reconnection in a fractal current sheet.

% Fig 14
\begin{figure}[t]
\sidecaption[t]
 \includegraphics[scale=.25]{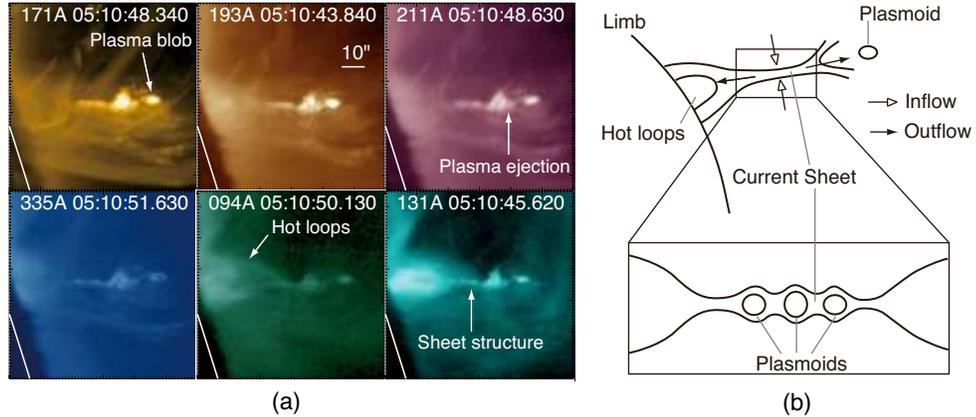}
  \caption{
 (a) Close-up images of the reconnection site of a solar flare in six different wavelengths (171, 193, 211, 335, 94, and 131 A) of AIA at the time when the current sheet, the plasma blob, and the hot post flare loops are observed. White solid lines indicate the solar limb.
(b) Schematic diagram of the flaring region. Black solid lines indicate the magnetic field. Top: the global configuration of the magnetic field. Bottom: a close-up image of the current sheet region. [From Takasao et al. (2012)]
}  
\label{fig:14}
\end{figure}

Takasao et al. \cite{Takasao2012} observed 
both reconnection inflow and outflow simultaneously
using SDO/AIA EUV images of a flare and
derived the nondimensional reconnection rate 0.055 - 0.2.
They also  found that during the rise phase of the flare, some plasma blobs appeared in the sheet structure above the hot flare loops, and 
they were ejected bidirectionally along the sheet
(see Fig. 14).
This is the first imaging observations of the plasmoid-dominated
current sheet in a solar flare. 

More recently, Nishizuka et al. (2015)\cite{Nishizuka2015} examined 
observational data of slowly drifting pulsating structures (DPSs) in the 
0.8 - 4.5 GHz frequency range taken with the radio spectrographs at 
Ondrejov Observatory.  
It is interesting to see that the DPSs are signatures of plasmoids, 
and from the observations of DPSs the plasmoid velocity and the reconnection rate were derived. 
The reconnection rate shows a good, positive correlation with the plasmoid velocity.
Nishizuka et al. (2015) also confirmed that some of the DPS events show plasmoid counterparts in
SDO/AIA images.

% section 5
\section{Stellar Flares}

\subsection{Unified Model of Solar and Stellar Flares: Emission Measure - Temperature Diagram}

The stellar flares show X-ray light curves similar to those of solar flares.
The time scale and typical properties derived from soft X-rays also show 
some similarities to solar flares, though dynamic range of stellar flare 
parameters are much wider than those of solar flares. Recent X-ray
astronomy satellites, such as ASCA, revealed that flares are frequently occurring 
in young stars, even in class I protostars 
(Koyama et al. 1996) \cite{Koyama1996}. 
One remarkable characteristics of these protostellar flares is that
the temperature is generally high, $50-100$MK, much hotter than the temperature
of solar flares, $10-20$MK. 
The total energy estimated is also huge, and amounts
to $10^{36-37}$ erg, much greater than that of solar flares, $10^{29-32}$ erg.

Can we explain the protostellar flares by magnetic reconnection models?
The answer is, of course, yes.  A part of the reason of this answer comes
from our finding of empirical correlation between emission measure and
temperature of solar, stellar, and protostellar flares. 
Figure 15 shows the observed relation between emission measure and
temperature of solar flares, microflares, stellar flares 
(Feldman et al. 1995)\cite{Feldman1995},  
and young stellar objects (YSO) flares \cite{Shibata-Yokoyama1999}. It is remarkable that these data
show the same tendency in a very wide dynamic range.  What does this relation 
mean ?

% fig 15
\begin{figure}[t]
\sidecaption[t]
 \includegraphics[scale=.7]{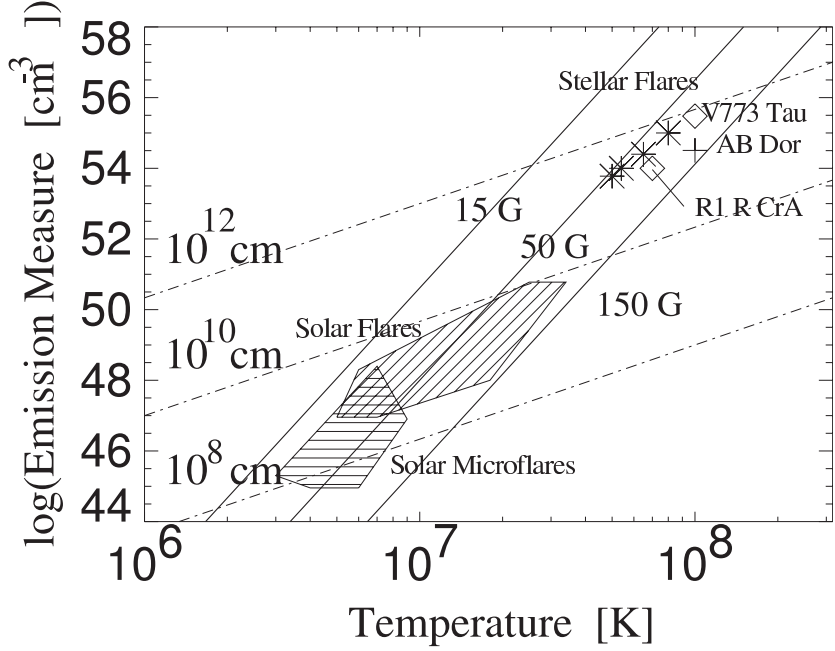}
  \caption{The EM (emission measure)$-$T (temperature) 
diagram for solar and stellar
flares and corona (Shibata and Yokoyama 2002)\cite{Shibata2002}.
Hatched area shows solar flares (oblique hatch) and solar microflares
(horizontal hatch), whereas other symbols denote stellar/protostellar flares.
Solid lines correspond to magnetic field strength = constant, and 
dash-dotted lines correspond to flare size = constant.}
\label{fig:15}
\end{figure}

Our answer is as follows (Shibata and Yokoyama 1999, 2002)
\cite{Shibata-Yokoyama1999},\cite{Shibata2002}. 
Yokoyama and Shibata \cite{Yokoyama1998},
\cite{Yokoyama2001} performed the self-consistent MHD
simulation of reconnection with heat conduction and evaporation for the first
time. From this simulation, they discovered a simple scaling relation for the
flare temperature:

%$$T \propto B^{6/7} L^{2/7}.    \eqno(1) $$
$$T \simeq 10^7  \Big({B \over 50 {\rm G}} \Big)^{6/7} 
       \Big({L \over 10^9 {\rm cm}} \Big)^{2/7}
       \Big({n_0 \over 10^9 {\rm cm}^{-3} } \Big)^{-1/7}   {\rm K}.  \eqno(8) $$

This is simply a result of energy balance between reconnection heating
($B^2 V_A/4\pi$) and conduction cooling ($ \kappa T^{7/2}/L$)
(since the radiative cooling time is much longer than the conduction time) .
With this equation and definition of emission measure ($EM = n^2 L^3$), and
pressure equilibrium ($p = 2nkT = B^2/8\pi $), we finally obtain the following
relation:

%$$ EM \propto B^{-5} T^{17/2}.     \eqno(2)  $$
$$ {\rm EM} \simeq 10^{48}  
      \Big({B \over 50 {\rm G}} \Big)^{-5} 
       \Big({T \over 10^7 {\rm K}} \Big)^{17/2} 
       \Big({n_0 \over 10^9 {\rm cm}^{-3}} \Big)^{3/2}
      {\rm cm}^{-3}.  \eqno(9) $$

We plotted this relation for constant field strength (B = 15, 50, 150 G)
in Figure 15. It is remarkable that these B = constant lines are consistent with the empirical correlation. In other words, the comparison between observation
and our theory tells that the magnetic field strength of solar and stellar flares
are not so different, of order of 50-150 G. In the solar case, this value
agrees well with the observations (average field strength of active region). 
In the case of stars, we have only limited set of observations,
but these observations show a kG field in the photosphere, suggesting a
100 G average field strength in the stellar corona, consistent with our 
theoretical prediction. 

We can also plot constant loop length lines in the diagram in
Figure 15. 

$$ {\rm EM} \simeq 10^{48}  
      \Big({L \over 10^9 {\rm cm}} \Big)^{5/3} 
       \Big({T \over 10^7 {\rm K}} \Big)^{8/3} 
       \Big({n_0 \over 10^9 {\rm cm}^{-3}} \Big)^{2/3}
      {\rm cm}^{-3}.  \eqno(10) $$

The loop length for microflares and flares is $10^8 - 10^{10}$ cm,
consistent with the observed sizes of microflares and flares, whereas 
the size of stellar flare loop is huge, even larger than $10^{11}$ cm, 
comparable to or even larger than stellar radius. Because of this large size,
the total energy of protostellar flares become huge and their temperature
becomes hotter than those of solar flares (see eq. 1). 
Since it is not possible to resolve the stellar flares, the large sizes of 
stellar flares  are simply theoretical prediction at present.

Shibata and Yokoyama (2002) \cite{Shibata2002} 
noted that the EM-T diagram is similar to
the Hertzsprung-Russell (HR) diagram, and examined basic properties of the EM-T diagram.
They found the existence of coronal branch, forbidden regions, and also
showed that flare evolution track can be plotted on the EM-T diagram, 
similarly to stellar evolution track in HR diagram.

\subsection{Superflares on Solar Type Stars}

It is well known that the first solar flare observed and recorded by human beings (Carrington 1859)\cite{Carrington1859}  was the largest solar flare ever observed and its released energy was estimated to be of order of $10^{32}$ erg (Tsurutani et al. 2003)\cite{Tsurutani2003}. This “Carrington flare" generated the largest geomagnetic storm in recent 200 years, and caused some damage to the telegraph system (Loomis 1861)\cite{Loomis1861} even in such a beginning phase of modern civilization based on electricity.  Is it possible for the Sun to produce “superflares" that are much more energetic than the “Carrington flare"? 

By analyzing existing previous astronomical data, Schaefer et al. (2000)\cite{Schaefer2000} discovered 9 superflares with energy $10^{33} \sim 10^{38}$ erg in ordinary solar type stars (G type main sequence stars with slow rotation with velocity less than 10 km/s). It was argued that the cause of the superflares is the hot Jupiter orbiting near to these stars (Rubenstein and Schaefer 2000)\cite{Rubenstein2000}, and thus concluded that the Sun has never produced superflares, because the Sun does not have a hot Jupiter (Schaefer et al. 2000).

Maehara et al. (2012)\cite{Maehara2012} analyzed the photometric data obtained by the Kepler space telescope (which was intended for detecting exoplanets using transit method), and found 365 superflares on 148 solar type stars.  Figure 16 shows a typical example of a superflare observed by Kepler, which shows the spike-like increase (1.5 percent) in stellar brightness for a short time (a few hous). It should be remembered that even one of the largest solar flares in recent 20 years (X18 class solar flare in 2003) showed only 0.03 percent solar brightness increase for 5 to 10 minutes. The total energy of this superflare was estimated to be around $10^{35}$ erg, 1000 times larger than the largest solar flare ($10^{32}$ erg).

% Fig 16
\begin{figure}[t]
\sidecaption[t]
 \includegraphics[scale=.8]{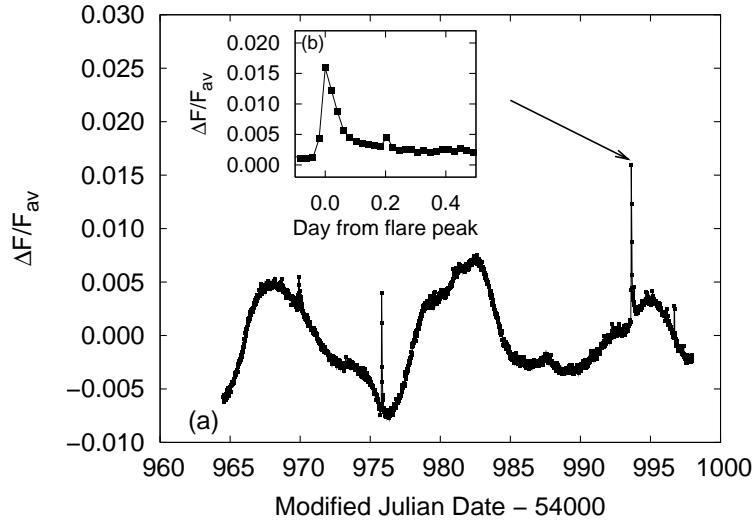}
  \caption{
A typical example of a superflare on a solar type star. (a) Light curve of superflares on the G-type main-sequence star KIC 9459362. (b) Enlarged light curve of a superflare observed at BJD2,454,993.63 (Maehara et al. 2012\cite{Maehara2012}).
 }
 \label{fig:16}
\end{figure}

It is also interesting to see in Figure 16 that the stellar brightness itself shows significant time variation with amplitude of a few percent with  characteristic time of 10 to 15 days. It is remarkable that almost all superflare stars show such a time variation of the stellar brightness. Maehara et al. (2012) \cite{Maehara2012} interpreted that the stellar brightness variation may be caused by the rotation of a star with big starspots.  Notsu et al. (2013)\cite{Notsu2013}  developed this idea in detail using the model calculation of the brightness change of the rotating star with big starspots. If this interpretation is correct, we can indirectly measure the rotation period of stars and the size of star spot (or total magnetic flux assuming the magnetic flux density is the same as that of the sunspot, 1000 to 3000 G). Since a big spot can store huge amount of magnetic energy around it, it is reasonable that almost all superflare stars show stellar brightness change of the order of a few percent or more. 

According to Shibata et al. (2013)\cite{Shibata2013}, the maximum energy of solar flares in a spot with magnetic flux density $B$ and an area $A$  has an upper limit determined by the total magnetic energy stored in a volume $A^{3/2}$ near the spot, i.e.,  

$$E_{flare} \simeq f E_{mag}
   \simeq  f {B^2 \over 8 \pi} A^{3/2}   
   \simeq 7 \times 10^{32} [{\rm erg}] 
   \Big( {f \over 0.1} \Big)
   \Big( {B \over 10^3 {\rm G}} \Big)^2
   \Big( {A  \over 3 \times 10^{19}  {\rm cm}^2}  \Big)^{3/2} $$  
$$
    \simeq 7 \times 10^{32} [{\rm erg}] 
   \Big( {f \over 0.1} \Big)
   \Big( {B \over 10^3 {\rm G}} \Big)^2
   \Big( {A/2 \pi {R_{sun}}^2  \over 0.001} \Big)^{3/2}  
  \eqno(11) $$
where $f$ is the fraction of magnetic energy that can be released
as flare energy.

Figure 17 shows the empirical correlation between the solar flare energy (assuming that GOES X-ray flux is in proportion to flare energy) versus sunspot area. We see that the theoretical relation (upper limit is used in eq. (11)) nicely explains observed upper limit of flare energy as a function of sunspot area. We also plotted the superflare data on the Figure 17. It is interesting to see that there exist many superflares above the theoretical upper limit. One possible solution of this apparent discrepancy is that these stars (above an upper limit) may be pole-on stars. Namely, if we observe stars from the pole, we tend to estimate smaller size of starspot, because the brightness change of stars (due to rotation) becomes small when viewing from rotating poles.  

Later, Notsu et al. (2015)\cite{Notsu2015}, using spectroscopic observations of 34 superflare stars, confirmed the interpretation, in addition to the confirmation of the real rotation velocity of these 34 stars (see also Notsu, S. et al. 2013\cite{NotsuS2013}, Nogami et al. 2014)\cite{Nogami2014}. 

Figure 17 shows that both solar and stellar flares are caused by the release of magnetic energy stored near spots. Figure 15 (EM-T diagram) along with Figure 17 (energy vs magnetic flux diagram) makes us sure that in a statistical sense the stellar flares are actually caused by the magnetic reconnection.

% Fig 17
\begin{figure}[t]
\sidecaption[t]
\includegraphics[scale=.8]{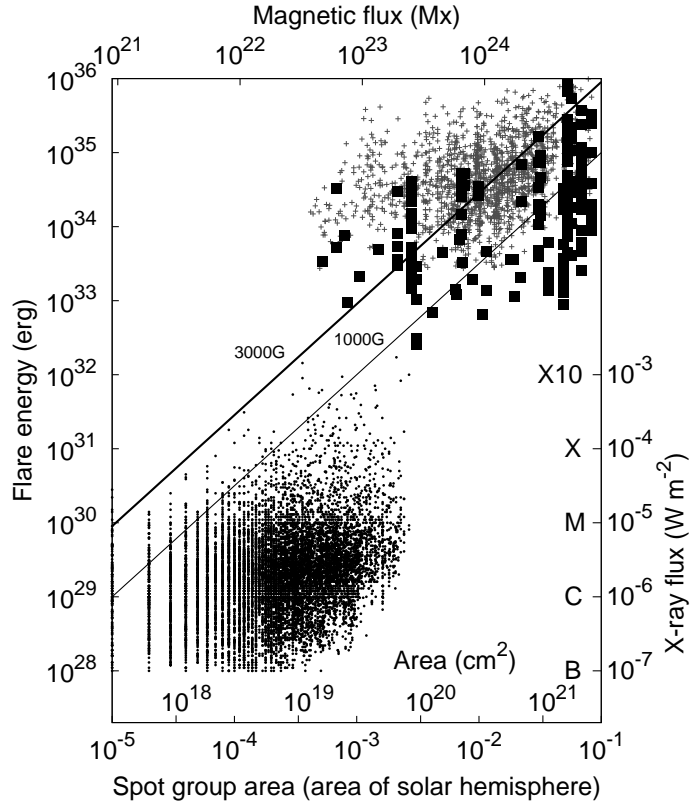}
  \caption{
%Flare energy vs sunspot area (\cite{Shibata2013})
Flare energy vs sunspot area (Maehara et al. 2015\cite{Maehara2015}).
Thick and thin solid lines in this figure represent Equation 11
for f = 0.1, B = 3, 000 and 1,000 G, respectively. Filled
squares and small crosses show data of superflares on solar type stars, 
while small dots are solar flare data \cite{Maehara2015}.  
}
\label{fig:17}
\end{figure}

Maehara et al. (2015)\cite{Maehara2015} analyzed the short time cadence data (1 min) taken by the Kepler mission, and found that the duration of superflares scales with flare energy ($E$) as $t_{flare} \propto E^{0.39}$, which is similar to the  correlation between the duration of solar flares and X-ray fluence $E$ observed with the GOES ($t_{flare} \propto E^{1/3}$) (Veronig et al. 2002\cite{Veronig2002}). This correlation is interesting because the reconnection model of flares predicts that the flare energy and duration scales with the length $E \propto L^3$ and $t_{flare} \propto L$, since the flare duration is basically determined by the inverse of the reconnection rate, of order of 100 $t_A = 100L/V_A$. From these relations, we find $t_{flare} \propto E^{1/3}$. This explains both solar and stellar flare observations. It provides another evidence of the magnetic reconnection model for {\it spatially unresolved} stellar flares.

What is the frequency of solar flares and stellar superflares ?   Figure 18 shows the  occurrence frequency of flares as a function of flare energy, for solar flares, microflares, nanoflares and also superflares on Sun-like stars.  It is remarkable to see that superflare frequency is roughly on the same line as that for solar flares, microflares, and nanoflares, 
$$ dN/dE \propto E^{-1.8}      \eqno(12)$$
suggesting the same physical mechanism for both solar and stellar flares. 
It was found that the occurrence frequency of superflares of $10^{34}$ erg is once in 800 years, and that of $10^{35}$ erg is once in 5000 years on Sun-like stars whose surface temperature and rotation are similar to those of the Sun.

It should be noted here that there is no evidence of hot Jupiters around the superflare stars, suggesting the possibility that superflares may occur on the Sun (Nogami et al. 2014\cite{Nogami2014}).  

Shibayama et al. (2013)\cite{Shibayama2013} extended and confirmed the work by Maehara et al. and found 1547 superflares on 279 solar type stars from 500 days Kepler data. 
Shibayama et al. found that in some Sun-like stars the occurrence rate of superflares was very high, four superflares in 500 days (i.e., once in 100 days).

% Fig 18
\begin{figure}[t]
\sidecaption[t]
 \includegraphics[scale=.8]{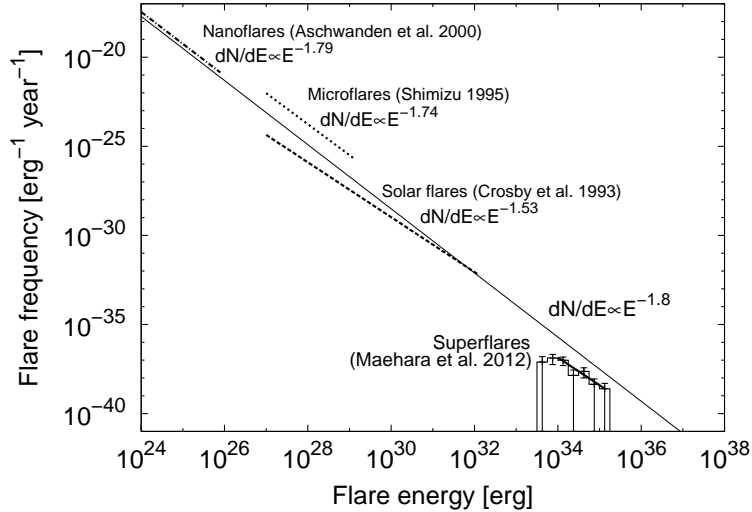}
  \caption{
Occurrence frequencies of solar flares, microflares, and nanoflares. 
Occurrence frequency of superflares on solar type stars are also shown  in this figure (Shibata et al. 2013\cite{Shibata2013})
}
\label{fig:18}
\end{figure}

It is interesting to note that large cosmic ray events in 7th and 9th century were found from tree ring (Miyake et al. 2012, 2013)\cite{Miyake2012},
\cite{Miyake2013}. 
Although the source of this cosmic ray is a matter of further investigation, the possibility that such event is caused by a solar super flare cannot be ignored. 
The frequency of the large cosmic ray events are pretty much consistent with the superflare frequency. 

If a superflare with energy $10^{34}-10^{35}$ erg 
(i.e., 100 - 1000 times larger than the largest
solar flares ever observed,  Carrington flare) occurs on the 
present Sun, the damage that such a superflare can cause to our civilization would be extremely large;
Hence it is very important to study the basic properties of superflare on
Sun-like stars to know the condition of occurrence of superflares and to understand how the superflares-producing stars are similar to our Sun. This is, of course, closely connected to the fundamental physics of reconnection: why and how fast reconnection occurs in magnetized plasma. 

Finally, we should note that stellar flares sometimes show very bursty 
light curves in X-rays and visibile light, which is similar to bursty radio or HXR light 
curves of solar flares during impulsive phase.  
This may be indirect evidence of turbulent (fractal) current sheet, 
since the fourier analysis of the time variability of the bursty light curve
shows a power-law distribution (e.g.,  Inglis et al. 2015\cite{Inglis2015}, 
Maehara 2015 private communication).

\begin{acknowledgement}
At first, we would like to thank Professor Eugene N. Parker for introducing us to the 
fascinating field of magnetic reconnection for many years.  
One of the authors (KS) remember that Prof. Parker said to KS
"What an interesting talk !"  just after KS gave a talk on "Plasmoid-induced-reconnection and fractal reconnection" in MR2000 conference
held in Tokyo in 2000. This comment encouraged KS very much, and 
it became the primary motivation why this article was written. 
We also would like to thank Amitava Bhattacharjee, Hantao Ji,  K. Daughton, N. F. Loureiro, Hiroyuki Maehara, Naoto Nishizuka, Yuta Notsu, Takuya Shibayama, Takuya Takahashi for allowing us to use figures of their papers and for their help for preparing the manuscript,  
Paul Cassak for giving us various useful comments as the referee, and
Alkendra Singh for improving our English. 
This work is supported by the Grant-in-Aids from the Ministry of
Education, Culture, Sports, Science and Technology of Japan (Nos.
25287039). ST acknowledges support by the Research Fellowship of the Japan Society for the Promotion of Science (JSPS). 
\end{acknowledgement}

\end{document}